\documentclass[reprint,onecolumn,superscriptaddress,a4paper,nofootinbib]{revtex4-1}
\usepackage[utf8]{inputenc}
\usepackage{graphicx}
\usepackage{footnote}
\usepackage{microtype}
\usepackage{amsmath}
\usepackage{amsfonts} 
\usepackage{color}
\usepackage{microtype}
\usepackage{graphicx}
\usepackage{footnote}
\usepackage{microtype}
\usepackage{amsmath}
\usepackage{amsfonts} 
\usepackage{amssymb}
\usepackage{textcomp}
\usepackage[section]{placeins}
\usepackage[font={small}]{caption}
\begin{document}
\title{Substrate induced proximity effect in superconducting niobium nanofilms}

\author{S.J. Rezvani}
\affiliation{SuperNanoLab, School of Science and Technology, Physics Division, University of Camerino, via Madonna delle Carceri 9, Camerino, 62032, Italy.}
\affiliation{Istituto Nazionale di Ricerca Metrologica (INRiM), Strada delle Cacce, 91, Torino, 10135, Italy}
\affiliation{IOM-CNR, Area Science Park, Trieste, Italy.}\footnote{Now at LNF-INFN, Via Enrico Fermi, Frascati, Rome, Italy}
\author{A. Perali}
\affiliation{SuperNanoLab, School of Pharmacy, Physics Unit, University of Camerino, via Madonna delle Carceri 9, Camerino, 62032, Italy.}
\author{Matteo Fretto}
\affiliation{Istituto Nazionale di Ricerca Metrologica (INRiM), Strada delle Cacce, 91, Torino, 10135, Italy}
\author{Natascia De Leo}
\affiliation{Istituto Nazionale di Ricerca Metrologica (INRiM), Strada delle Cacce, 91, Torino, 10135, Italy}
\author{L. Flammia}
\affiliation{SuperNanoLab, School of Science and Technology, Physics Division, University of Camerino, via Madonna delle Carceri 9, Camerino, 62032, Italy.}
\author{Milorad V. Milo\v{s}evi\'{c}}
\affiliation{Department of Physics, University of Antwerp, Groenenborgerlaan 171, B-2020 Antwerp, Belgium}
\author{S. Nannarone}
\affiliation{IOM-CNR, Area Science Park, Trieste, Italy.}
\author{N. Pinto}
\affiliation{uperNanoLab, School of Science and Technology, Physics Division, University of Camerino, via Madonna delle Carceri 9, Camerino, 62032, Italy.}





\begin{abstract}
Structural and superconducting properties of high quality Niobium nanofilms with different thicknesses are investigated on silicon oxide and sapphire substrates. The role played by the different substrates and the superconducting properties of the Nb films are discussed based on the defectivity of the films and on the presence of an interfacial oxide layer between the Nb film and the substrate. The X-ray absorption spectroscopy is employed to uncover the structure of the interfacial layer. We show that this interfacial layer leads to a strong proximity effect, specially in films deposited on a SiO$_2$ substrate, altering the superconducting properties of the Nb films. Our results establish that the critical temperature is determined by an interplay between quantum-size effects, due to the reduction of the Nb film thicknesses, and proximity effects.
\end{abstract}
\maketitle

\section{Introduction}

Superconductivity has been recently shown to survive even in extremely confined nanostructures such as metal monolayers \cite{Zhang2010}. Preserving a
superconducting state in ultrathin films can be achieved by nanofabrication techniques and withstand multiple cooling cycles. However, control over the superconducting properties of metallic ultrathin films are of utmost importance in realization of quantum devices, such as Josephson junctions, nano Superconducting QUantum Interference Devices (SQUIDs), mixers and single photon detectors \cite{Semenov:2003,karasik}. 
Hence, a systematic investigation on their properties and their optimization at reduced dimensionality can leads to an ideal platform to generate novel phenomena such as multi-gap and resonant phenomena in a controlled and optimized mode \cite{Blatt1,Blatt2,Innocenti2010,PeraliSUST,2DEGPerali,MMAP2015}. 

Niobium is an ubiquitous material for superconducting thin films and its performance is known to increase when epitaxy conditions can be reached \cite{Oya}. On the other hand, the proximity effect occurs when a superconductor is placed in contact with non superconducting materials. The resulting critical temperature of the superconductor in this case is suppressed and signs of weak superconductivity are induced in non superconducting materials. Superconducting correlations are induced in the normal side up to a distance where the electron and hole lose phase-coherence \cite{PG2013}. 
However, a combined electrical and structural study on the presence and role of the proximity effect induced by the substrate on the superconducting properties of Nb ultra-thin
films is lacking. 

In this work we report an extended experimental investigation of the superconducting to normal state transition in Nb nanofilms, with a thickness in the range 9 nm to  80 nm, deposited on SiO$_2$ and Al$_ 2$O$_3$ substrates.
A detailed analysis of several superconducting properties of these Nb nanofilms has been reported in \cite{Pinto2018}.
The main microscopic parameters characterizing the normal and the superconducting state of these films were investigated based on recent models. We have explored the important role played by the substrate on the fundamental properties of the Nb nanofilms. Our study establishes the existence of an interplay between quantum size and proximity effects at the substrate interface \cite{Chen2012,Palestini2013}.

\section{Experiment}
\begin{figure}
\includegraphics[width=\linewidth]{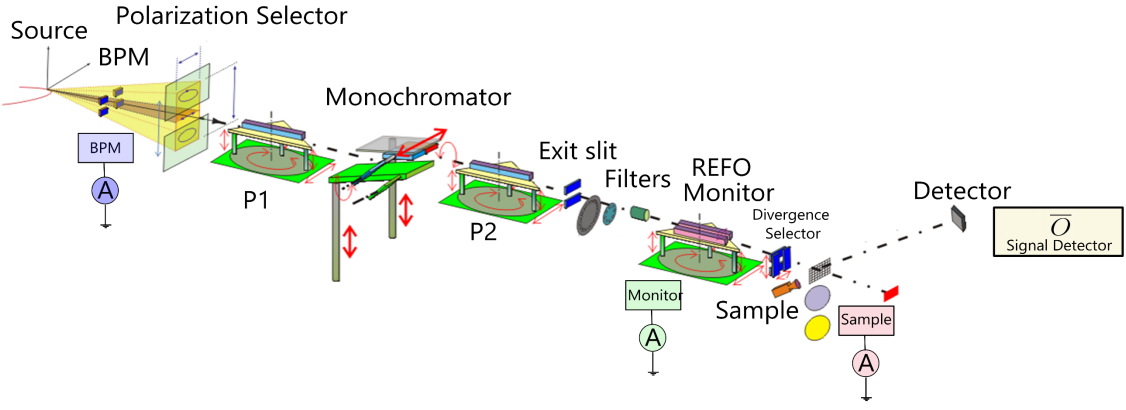}
\caption{BEAR beamline set up \cite{bear3}.} 
\label{bear}
 \end{figure}

Nanofilms of Niobium have been deposited at room temperature on thermally oxidized Si wafer (silicon oxide thickness: $300\div500$ nm) and on sapphire, in an ultra high vacuum dc sputtering chamber, with a base pressure $\approx 2\times 10^{-9}$ mbar. Films were deposited with thicknesses in the range from 9 nm to 80 nm with a constant deposition rate of 0.65 nm/s. Scanning electron microscopy (SEM) analysis has been carried out on some films by a FEI Quanta$^{TM}$ 3D FIB (Nanofacility Piemonte, INRiM).
Electronic transport properties of samples were measured using a Hall bar geometry, $1 \div 2$ cm long, 10 $\mu$m and 50 $\mu$m wide (see figure \ref{rho}a).
The resistivity, $\rho(T)$, was measured as a function of the temperature, in the range 4 $\div$ 300 K, by a He closed cycle cryostat  (ARS mod. DE-201S) equipped with two silicon diode thermometers (Lakeshore mod. DT-670). Resistivity has been measured sourcing a constant current (Keithley mod. 220), monitored by a pico-ammeter (Keithley mod. 6487) and a multimeter measuring the voltage drop (Keithley mod. 2000). The current was sourced in the range $1 \div 50$ $\mu$A. Further details can be found in the reference \cite{Pinto2018}.
The X-ray absorption measurements were carried out on oxygen K-edge on a 80 nm thick Nb film at IOM-CNR, BEAR beam-line at Elettra synchrotron radiation center (Trieste, Italy) \cite{bear1,bear2}. The beamline operates in the 2.8-1600 eV (443-0.775 nm) spectral region, delivering polarized light of selectable ellipticity from nearly linear to elliptical (see figure \ref{bear}). 

The samples were mounted in the experimental chamber after the divergence selector slits with a base pressure of 1$\times 10^{-9}$ and were positioned at 45$^\circ$ with respect to the incident beam (With a solid angle of $\sim 0.8$ strad). A reference current was measured simultaneously to the measurement to normalize the measured spectra to the variation of the photon intensities, while the incident beam was measured after removal of the sample from the experimental chamber.  The measurements were carried out in total fluorescence mode to achieve a mean probing depth of $\sim100$ nm \cite{Rezvani2017} to include both Nb film and substrate interface. The overall error of measurement on fluorescence yield was estimated of the order of 1\%.  

\section{Results and discussion}
\begin{figure}[h!]
\centering
\includegraphics[width=0.8\linewidth]{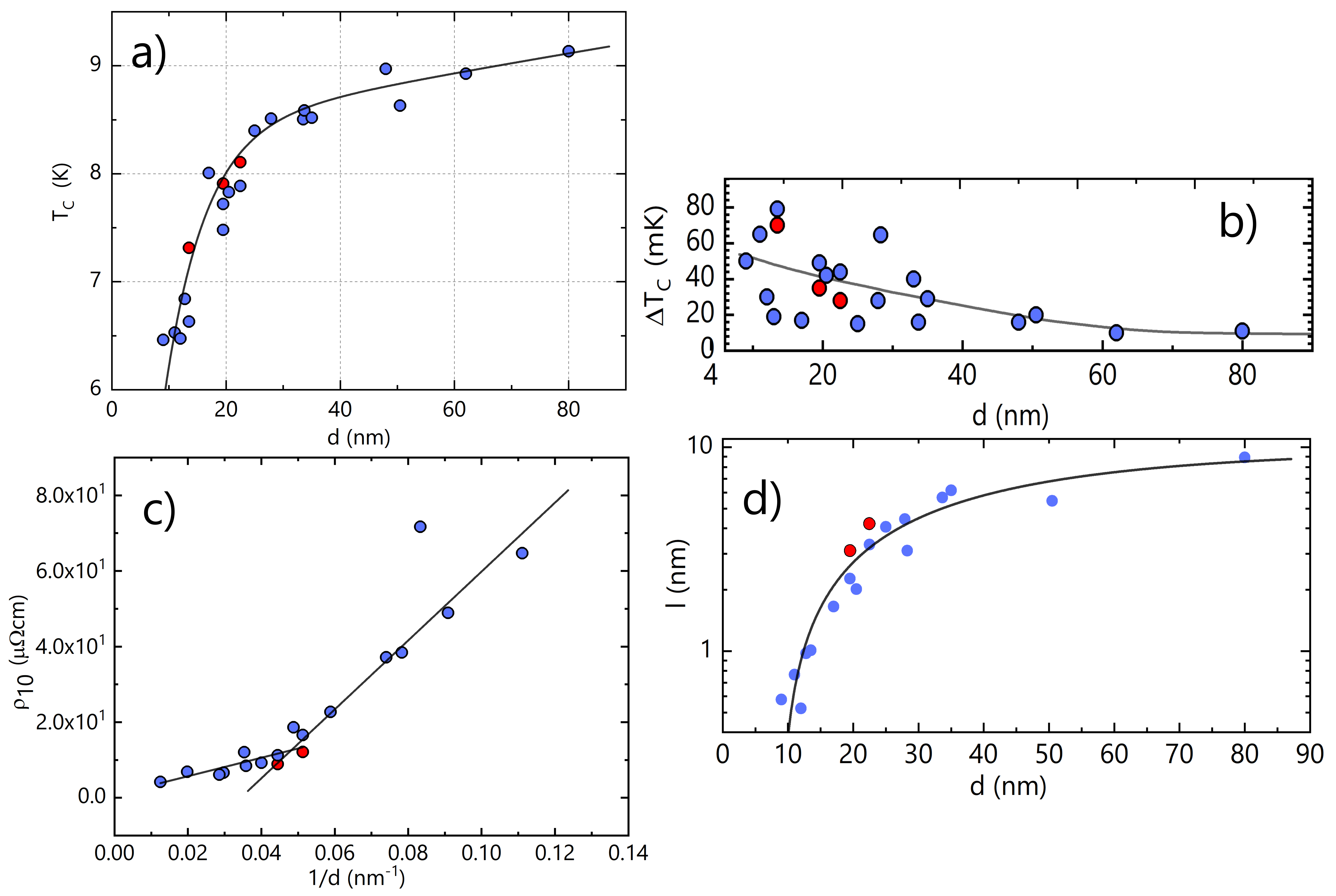}
\caption{Thickness dependence of a) T$_c$ on the thickness of Nb films deposited on SiO$_2$ (blue) and sapphire (red); b) The superconducting transition width. Red circles are Nb Films on sapphire. Continuous line is a guide for eyes; c) The residual resistivity for films on SiO$_2$ and sapphire; and d) The mean free path at 10 K. Data have been derived from the resistivity values measured at the same temperature (i.e. $\rho_{10}$).} 
\label{rho}
\end{figure}

The resistivity of the Nb films $\rho$ and its temperature dependence was investigated as a function of the film thickness, $d$ (see figure \ref{rho}a). 
Results show that the superconducting transition temperature, $T_C$, decreases quite abruptly as $d$ is progressively reduced while the $\rho(T)$ curves shift upwards. 
At a thickness equal to 80 nm our films resistivity approaches the expected Nb bulk value of 15 $\mu\Omega$cm \cite{Bulk_res1,Bulk_res2}. The decrease of the film resistivity in thicker films may suggest a gradual reduction of the film defectivity via atomic rearrangement \cite{Zhao,Delacour,Gubin2005}.  

Compared to the data reported in the literature, the $T_C$ of Nb films deposited on silicon oxide show lower values for smaller thicknesses. On the other hand, while in general $T_C$ shows a strong decreasing behavior as a function of film thickness, Nb films deposited on sapphire consistently demonstrate higher $T_C$ compared with the ones deposited on SiO$_2$ (Fig. \ref{rho}a). The residual resistivity (i.e. $\rho_{10}$) at $T = 10$ K (see figure \ref{rho}c) of films with different thicknesses also show an increasing trend by reduction of the films thickness while, this increase is less pronounced in films deposited on sapphires.   
The charge carriers mean-free path $l$, was estimated from the residual resistivity (i.e. $\rho_{10}$) at $T = 10$ K, ranging from $\approx 1$ nm, at the lower thicknesses, to $\simeq 9$ nm at $d = 80$ nm (Figure \ref{rho}d). The results show also a higher mean free path ($l$ values) for the Nb films deposited on sapphire. The width of the superconducting transition, $\Delta T_C$, also exhibits an increasing trend from 15 mK at 80 nm to 80 mK for 10 nm films (Figure \ref{rho}b). While $\Delta T_C$  in our samples show lower values compared to those reported in the literature, suggesting a relatively higher quality films \cite{Zhao}, the transition width of the Nb films on sapphire are relatively lower.

These results suggest an enhancement of the superconducting properties on sapphire substrate that can be associated with two distinct effects. First, the films deposited on the sapphire are less defective since sapphire lattice parameter and thermal dilatation coefficients match rather well with Nb \cite{Oya}. Second, this effect can be related to a contribution from an oxidized Nb (i.e. NbO$_x$) layer \cite{Gershenzon}, formed at the interface film-SiO$_2$ substrate, becoming progressively stronger with the reduction of $d$. This oxide layer reduces the effective Nb film thickness leading to a lower $T_C$. Such a contribution is largely reduced on sapphire substrates as can be seen in figure \ref{rho}d. Moreover, NbO$_x$ layer, being a conductive system of electrons in its normal state, can sink Cooper pairs from the superconducting Nb nanofilm, suppressing the condensate fraction and reducing the $T_C$ via proximity effect. This phenomenon dominates as the thickness of the Nb film is reduced with the NbO$_x$ layer becoming a sizeable fraction of the Nb film thickness, similar to the behaviour of $T_C$ in our films. 
The formation of Nb nanofilms with lower defectivity on the sapphire substrate is confirmed by the higher $l$ values on sapphire indicating a lower defect density in the film matrix.

\begin{figure}[ht]
\centering
\includegraphics[width=\linewidth]{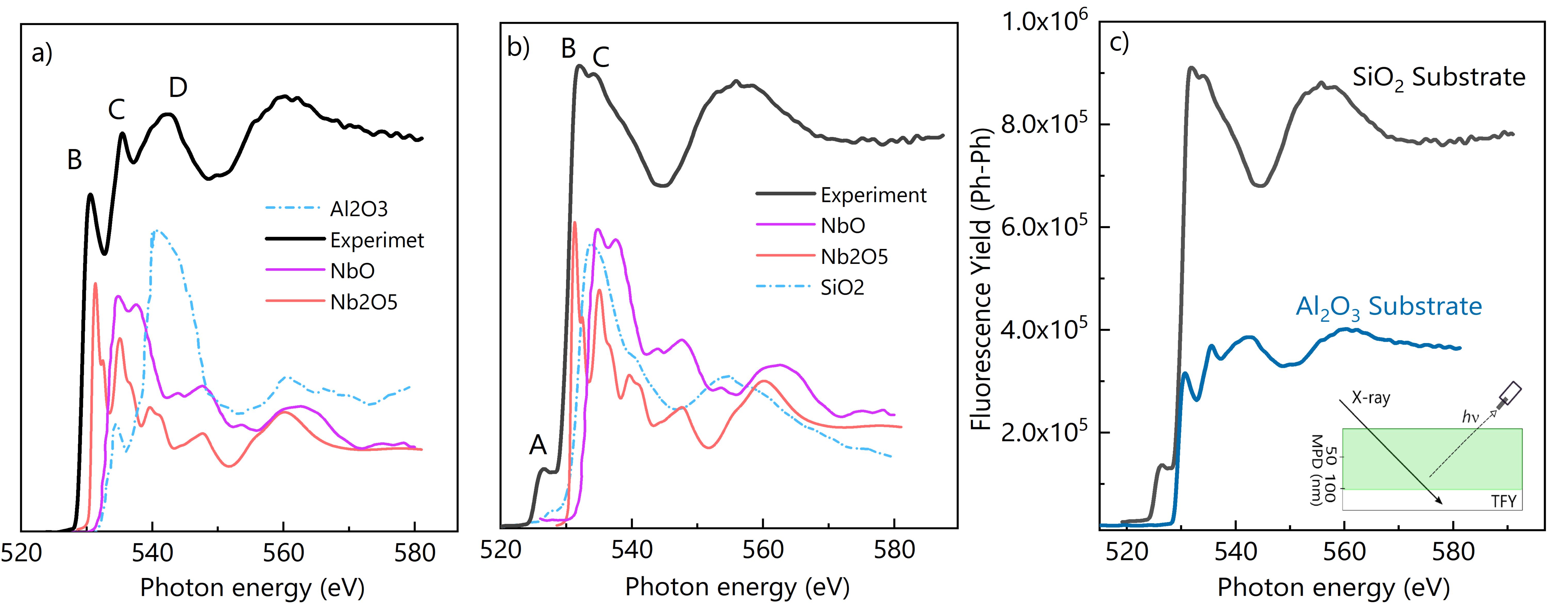}
\caption{X-ray absorption spectra of the O K-edge of the films deposited on the a) sapphire and b) silicon oxide in total fluorescence yield mode along with the reference spectra of SiO$_2$, Al$_2$O$_3$, Nb$_2$O$_5$ and NbO. c) normalized florescence yield, measured on samples on SiO$_2$ and sapphire along with the MPD of the TFY schematic. The overall yield error was estimated of the order of 1\%.} 
\label{xas}
\end{figure}

In order to investigate the formation of an oxide layers, at Nb surface and its interface with substrate, we carried out detailed experiments by X-ray absorption measurements in total florescence yield mode (TFY). This technique, with a mean probing depth (MPD) of $\sim100$ nm is perfectly suitable to reach the interface layer of a 80 nm thick Nb film \cite{Rezvani2017}. The TFY spectra of the films on silicon oxide and sapphire at oxygen K-edge are reported in figure \ref{xas}a,b along with the reference spectra of SiO$_2$, Al$_2$O$_3$, Nb$_2$O$_5$ and NbO. As shown clearly in the figure, samples on both substrates show two main features at around 531 and 535 eV (B and C) corresponding to the stable Nb$_2$O$_5$ phase (B), while for the sample on SiO$_2$ the component C is broadened due to the signal from the substrate itself. Furthermore, the increased intensity of the component C of the sample on sapphire and the broad component (D) as well, can be associated to the contribution of the sapphire substrate. These results confirm the formation of a major Nb$_2$O$_5$ phase on both samples including the superficial layer. 
However, the TFY spectra of the Nb/SiO$_2$ film exhibits an extra pre-edge structure (A) that can not be corroborated neither with usual niobium oxides phases, such as Nb$_2$O$_5$ or NbO, nor with the SiO$_2$ substrate itself and hence, signifies formation of a semi metallic Nb oxide (e.g., NbO$_x$). 
This semi metallic layer may induce a strong proximity effect, potentially sinking Cooper pairs from the superconducting Nb nanofilms and then leading to a strong reduction of the condensate fraction. Moreover, it can be seen that the normalized signal of the Nb/SiO$_2$ sample (figure \ref{xas}c) shows significantly higher intensity compared to that of Nb/Al$_2$O$_3$. This indicates higher concentration of the oxygen absorbers in case of Nb/SiO$_2$ film that can be associated to the formation of a larger oxide film. The larger oxide layer can reduce the effective thickness of the superconducting Nb film resulting in a further alteration of the superconducting properties. A detailed analysis of the formation of the interfacial and superficial oxides layers on the superconducting Nb films is the object of a forth coming paper Ref.\cite{future}.   

\begin{figure}[ht]
\centering
\includegraphics[width=0.5\linewidth]{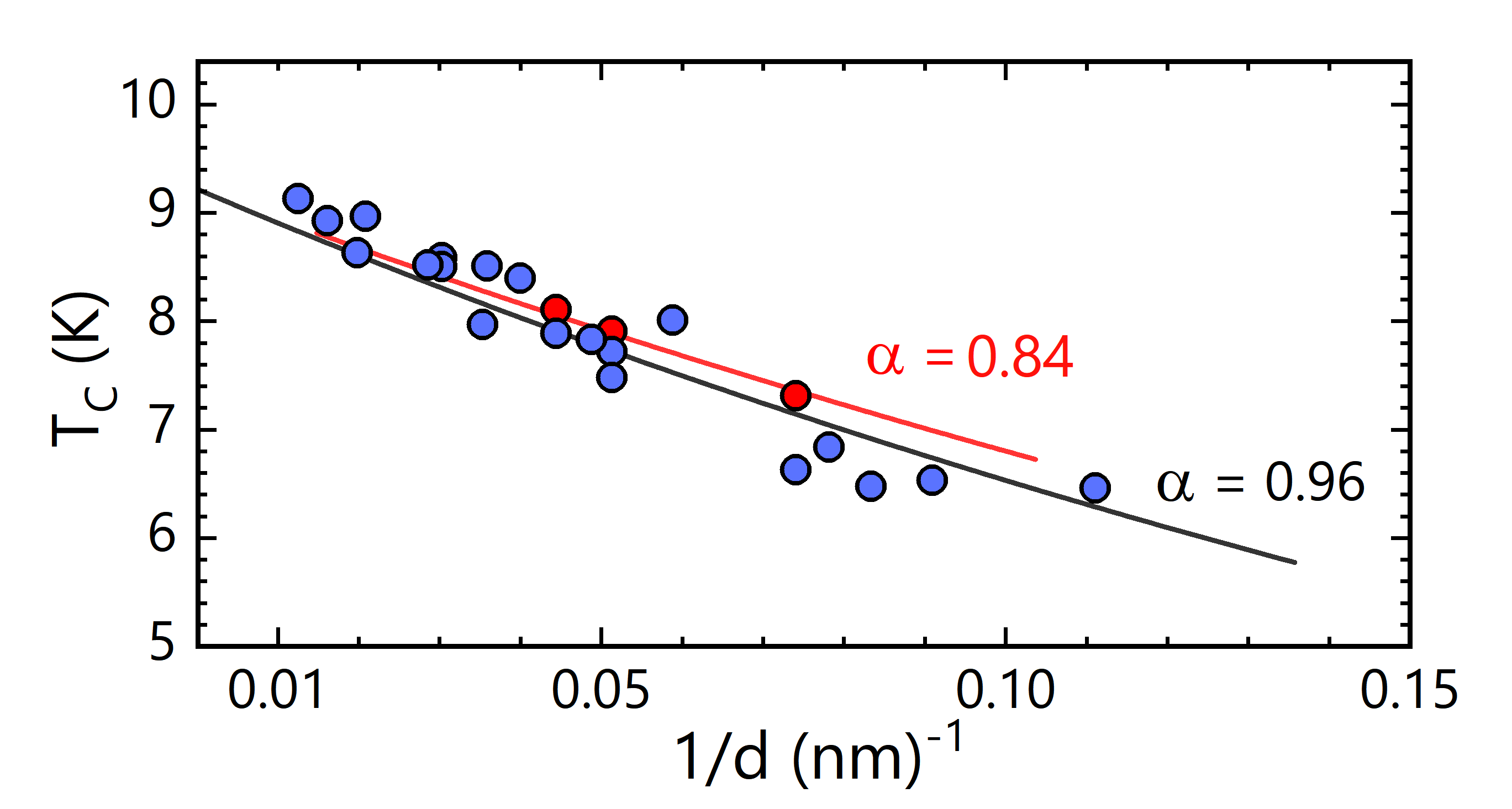}
 \caption{Top) Superconducting transition temperature as a function of the reciprocal film thickness. Blue circles are Nb films on SiO$_2$ substrate and red circles on sapphire substrate. Continuous line is least-squares fit, using Eq. \eqref{millan}. The effective thickness of the normal layer (in nm) is represented by the corresponding $\alpha$ value. Bottom) Plot of the ratio $T_c/F(d)$ as a function of the film thickness. The fitting function, $F(d)$, has been calculated using the theoretically expected value of the Nb bulk transition temperature (i.e. $T_{C0} = 9.22$ K).} 
\label{prox}
\end{figure}

Finally, to investigate the $T_C$ suppression due to the proximity effect induced by superficial/interfacial oxide layers, an approach by McMillan\cite{McMillan} was employed (Equation \ref{millan}). In this approach $\alpha = d_N N_N(0)/N_S(0)$ is an effective thickness of the conductive layer at the interface; $T_{C0} = 9.22$ K is the bulk $T_c$ of Nb and $T_D$ = 277 K is the Debye temperature. The quantities $N_N(0)$ and $N_S(0)$ are the density of states in the normal ($N$) and superconducting ($S$) layers, respectively and here assumed to be equal \cite{Pinto2018}.
\begin{equation}
\centering
T_C = T_{C0}\left( \frac{3.56 T_D}{T_{C0}\pi} \right)^{-\alpha/d}
\label{millan}
\end{equation}  
Fitting our data using equation \ref{millan} for $T_c(d)$ we obtain a value of $\alpha = 0.96\pm 0.04$ nm (Figure \ref{prox}a) for films on SiO$_2$ and $\alpha = 0.84\pm 0.02$ nm for films on sapphire. However, the $\alpha$ value for the films deposited on sapphire are calculated from few points available in our experiment.  These results are in a good agreement with the TFY results and can reasonably correspond to two different thin oxide layers induced by the substrates, resulting in a thinner oxide layer when Nb films are deposited on sapphire. 
Furthermore, in the presence of an overall suppression of $T_c(d)$ due to proximity effect, the fitting function given by Eq. \eqref{millan} can be subtracted from the data set in order to amplify the visibility of probable oscillations of $T_c$ for decreasing $d$ due to the incipient quantum size effects and shape resonances associated with the electronic confinement in the perpendicular direction. Subtracting the fit from the experimental data, we observe progressively increasing residual $T_c$ oscillations by decreasing $d$, with amplitude of ~5\% in the thinnest films, as reported in figure 2b, comparable to the theoretical predictions for Pb and Al nanofilms \cite{Shanenko2006B}.

\section*{Conclusions}
In this work we have investigated the effects of the substrate (silicon oxide and sapphire) on the superconducting properties of thin Nb films via structural, electrical and superconducting analysis.
While our results demonstrate generally high quality nanofilms, nonetheless Nb films on sapphire substrates consistently exhibit improved superconducting properties, compared with those on SiO$_2$ of the same thickness. It is shown that this improvement can be the result of different effects such as the lower defectivity of Nb deposited on sapphire and formation of a semi metallic oxide layer at the interface with the SiO$_2$ substrate, observed by X-ray absorption spectroscopy. The XAS results also indicate a higher concentration of oxide in Nb films on SiO$_2$ substrates that may suggest a higher thickness of oxide layers. The higher thickness of the oxidized layers may leads to a lower effective thickness of the Nb films altering their superconducting properties. Our results point towards a significant contribution of proximity effect in the superconducting properties of the films, particularly for the films on SiO$_2$. 

\section*{Acknowledgments}
The authors would like to thank A. Giglia and K. Koshmak for their assistance in X-ray Absorption Spectroscopy measurements and L. Pasquali for constructive discussions.  

\bibliography{rezvani_arxiv}

\end{document}